\documentclass[11pt,twoside]{article}

\usepackage{asp2006}
\usepackage{epsf}
\usepackage{epsfig}
\usepackage{lscape}
\usepackage{natbib}

\newcommand{\emax}{e_{\rm max}}

\newcommand{\Mj}{M_{\rm J}}
\newcommand{\aB}{a_{\rm B}}
\newcommand{\eB}{e_{\rm B}}
\newcommand{\IB}{I_{\rm B}}
\newcommand{\Msol}{M_\odot}

\begin{document}
\title{Secular Evolution of Planetary Systems in Binaries}   
\author{Genya Takeda, Ryosuke Kita, Frederic A. Rasio and Samuel M. Rubinstein}   
\affil{Department of Physics and Astronomy, Northwestern University, 2145 Sheridan Rd., Evanston, IL 60208-3112 USA}    

\begin{abstract} The orbital eccentricity of a single planet in a stellar binary system with a sufficiently large mutual inclination angle is known to oscillate on a secular timescale through the Kozai mechanism.  We have investigated the effects of the Kozai mechanism on double-planet systems in binaries.  The evolutionary sequence of a pair of planets under the influence of a binary companion is fairly complex.  Various dynamical outcomes are seen in numerical simulations.  One interesting outcome is the rigid rotation of the planetary orbits in which the planetary orbital planes secularly precess in concert, while the orbital eccentricities oscillate synchronously.  In such cases the outer planet acts as a propagator of the perturbation from the binary companion to the inner planet and drives the inner planetary orbit to precess at a rate faster than what is predicted by the Kozai mechanism.
\end{abstract}



\section{The Kozai Mechanism \label{intro}}
The first decade of radial-velocity and photometric surveys has revealed that planetary systems commonly form among stellar binary systems.  As of July 2007, at least 37 planetary systems ($\sim 20\%$ of all the known systems) are members of multiple-stellar systems \citep{raghavan06,desidera07}.  The true stellar multiplicity among planetary systems may be higher than $20 \%$ since the photometric surveys in search for stellar companions around currently known planetary systems are not yet complete.  Because most of these stellar companions are in wide orbits (typically $a = 10^2$\,AU or greater), their orbital inclinations are not correlated with the invariable plane of the planetary systems, implying that many binary companions orbit around planetary systems in largely inclined orbits \footnote{If the distribution of the mutual inclination angle is completely isotropic, the probability of the mutual inclination angle being greater than the critical Kozai angle ($39.2\deg$) is $77\%$}.  

\begin{figure}[!ht]
\centerline{\includegraphics[scale=.4]{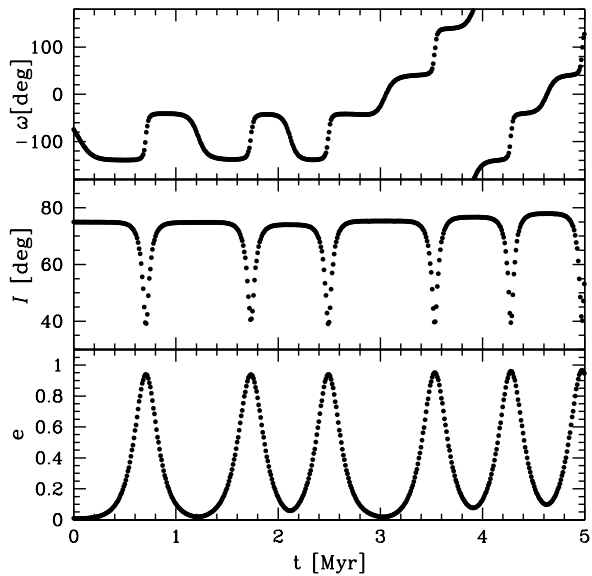}}
\caption{An example of secular  evolution of a planet orbit through the Kozai mechanism.  The amplitude of the eccentricity oscillation ($\emax$) is determined by the initial relative inclination angle.  The argument of periastron typically alternates between libration and circulation during the Kozai cycles. \label{sample_kozai}}
\end{figure}

It has been known in celestial mechanics that in a hierarchical triple-body system with  sufficiently large mutual inclination angle ($I > I_{\rm crit} = 39.2^\circ$) between the inner and outer binaries, the inner binary's orbital eccentricity and inclination undergo large-amplitude oscillations on a secular timescale through the ``Kozai mechanism'' \citep{kozai62}.  When adopted to planetary systems, the inner binary orbit can be replaced by the osculating orbit of a planet around the primary star.  The Kozai mechanism has following unique properties (see also Figure~\ref{sample_kozai}): (1) the orbital eccentricity of the planet oscillates with an amplitude that is  dependent (in the lowest order) only on the initial mutual inclination angle $I_0$; $\emax = \sqrt{1 - 5/3 \cos{I_0}^2}$.  Other orbital parameters only affect the eccentricity oscillation period, 
\begin{equation}
P_{\rm e} \simeq P_{\rm b}^2 / P_{\rm pl}\, (1 - e_b^2)^{3/2} ,  \label{Pkoz}
\end{equation}
 \citep{ford00} where $P_{\rm pl}$ and $P_{\rm b}$ are the orbital periods of the planet and the binary, respectively, and $e_{\rm b}$ is the binary eccentricity.
(2) The orbital inclination $I$ of the planet with respect to the binary plane also oscillates on the same timescale as the eccentricity oscillation, conserving the Kozai integral $H_{\rm K} = \sqrt{1 - e^2} \cos{I}$.  (3) The argument of the periastron $\omega$ alternates between the libration mode around $\pm 90^\circ$ and the circulation mode \citep{fabrycky07}.

An episode of the Kozai mechanism in an extrasolar system may leave a directly observable signature in the orbital eccentricity of the planet.  Theoretical models show that several highest planetary eccentricities observed are best explained by the Kozai mechanism \citep{holman97,wu03}.  Also, the median eccentricity of the planets in binaries ($e_{\rm med,binary} = 0.25$) is higher than that of the planets around single stars ($e_{\rm med, single} = 0.20$), implying that the planetary eccentricities are commonly excited by stellar companions.  

The Kozai eccentricity oscillation results from the torque exerted by the binary companion averaged over the planetary orbit.  If there is another secular perturbation at play that precesses the planetary orbit in a shorter timescale than the Kozai oscillation timescale (Eq.~\ref{Pkoz}), the averaged Kozai torque is reduced, and the Kozai eccentricity oscillation will be suppressed.  Extra precessions are typically caused by general relativistic effects, tidal torques, or mutual interactions between planets \citep{wu03,takeda05}.

\section{Evolution of Double-planet Systems in Binaries}
\subsection{The Classical Secular Theory   Multiple-Planet Systems\label{LL}}
Although the Kozai mechanism has been successful in modeling single planets in binaries, there are only handful of literatures that investigated the evolution of multiple-planet systems in binaries.  What makes the problem challenging is the large dimensions of the parameter space; six orbital elements and the mass for the two planets and the binary companion already sum up to 21 parameters, virtually prohibiting a simple numerical exploration of the entire parameter space.  Analytical understanding is necessary to identify different evolutionary outcomes. 

The extra complexity for the evolution of double-planet systems in binaries is due to the additional secular torque which planets exert on each other in addition to the Kozai torque applied by the binary companion.  All the torques result in the secular precession of planetary orbits and evolution of orbital eccentricities.  The secular orbital solutions of  planetary systems through mutual planet-planet interactions  have been formulated through the classical Laplace-Lagrange theory (L-L theory, hereafter).  Assuming small initial orbital eccentricities and inclinations \footnote{Typically, the terms up to the second order in $e$ and $I$ are included in the standard L-L theory.  The generalized fourth-order L-L theory has been recently developed by \citet{veras07}.}, the L-L theory can predict from initial conditions the evolutionary sequence of the orbital parameters, namely $e(t), \omega(t), I(t), $ and $\Omega(t)$.   It should be kept in mind that the L-L theory is an approximated perturbation theory and is not a reliable estimator for predicting accurate long-term behavior of a planetary system.  However, the great merit of the L-L theory is that one can extract useful evolutionary information through simple linear algebra and derive various order-of-magnitude estimates for evolutionary timescales.

\subsection{Numerical Simulations}
We have run a large set of numerical simulations with various initial conditions to investigate  different evolutionary outcomes of hierarchical four-body systems (i.e., double planets orbiting a component of a wide stellar binary).    The equations of motion were fully integrated by the Bulirsch-Stoer method to account for all the possible close encounters between planets or between a planet and the primary star.  To reduce the number of free parameters, we have used fixed binary parameters: $\aB = 750\,$AU, $\IB = 50\deg$ (with respect to the invariable plane of the planets), $\eB = 0.2$, and both binary components with masses $M_0 = M_{\rm B} = 1 \Msol$.  Also, initially the two planets are both in near-circular orbits with eccentricities less than 0.01 and separated by more than one mutual Hill radius to ensure that the system does not immediately become unstable.  All the systems were evolved for 500\,Myr, a time sufficient for the outer planet to undergo at least a few Kozai cycles, unless the system has gone unstable.  

\begin{figure}[!t]
\centerline{\includegraphics[scale=.45]{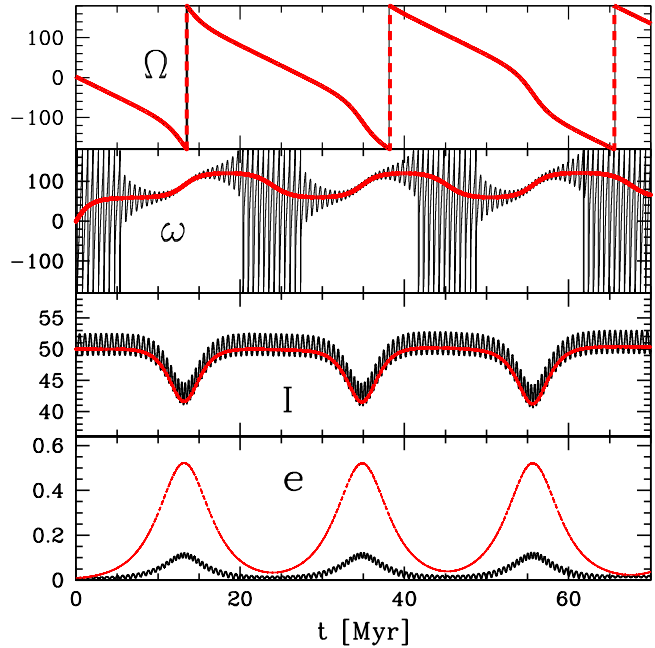}}
\caption{A numerical simulated evolution of rigidly rotating double-planet system.  Dashed curves represent the orbital elements of the outer planet.  All the angular quantities are denoted in degrees.  The longitudes of ascending node  and the orbital inclinations of the two planets are almost indistinguishable throughout the evolution.  The orbital eccentricities of both planets oscillate at the same period.  Notice that when $e_2$ is excited to $\ga 0.1$ by the Kozai mechanism, the periastron arguments get locked into tight apsidal libration. \label{sample_rigid}} 
\end{figure}

\subsection{Rigid Rotation of Planetary Systems}
A variety of dynamical outcomes have been observed in our simulations.  Of particular interest was the rigid rotation of the planetary system, in which the orbital elements of both planets secularly evolve in concert.  A similar result has been previously observed by \citet{innanen97}.  They have numerically calculated the evolution of the Solar planets with the presence of a hypothetical stellar companion to the Sun and observed that the inclinations and ascending nodes of the planets are always aligned throughout the evolution, as if the planets had been embedded in a rigid disk.

Figure~\ref{sample_rigid} shows an example of a rigidly rotating double-planet system.  It is immediately noticeable that the orbital inclinations and the longitudes of ascending node are strongly aligned throughout the evolution, maintaining mechanical rigidity of the orbital planes.  Another interesting aspect of Figure~\ref{sample_rigid} is that $e_1$ oscillates at the same periodicity as $e_2$, but with a smaller amplitude.  These synchronous eccentricity oscillations are contrary to what would be predicted by the Kozai mechanism; a pair of coplanar planets at different orbital radii are expected to undergo  eccentricity oscillations at equal amplitudes but with different periods.  

The synchronized eccentricity oscillations result from the small-amplitude apsidal libration, for which $\Delta \omega  \equiv \left| \omega_1 - \omega_2 \right| < \Phi$ is maintained for a small angle $\Phi$.   It has been previously discovered by \citet{chiang02} that the adiabatic increase of the outer planet's eccentricity due to the remnant gas disk drives the apses of the planets from independent precessions (circulation) to strongly coupled precessions (aligned libration).  The same apsidal trapping mechanism is essentially what causes the Kozai-induced synchronized eccentricity oscillations. 

\begin{figure}[!t]
\centerline{\includegraphics{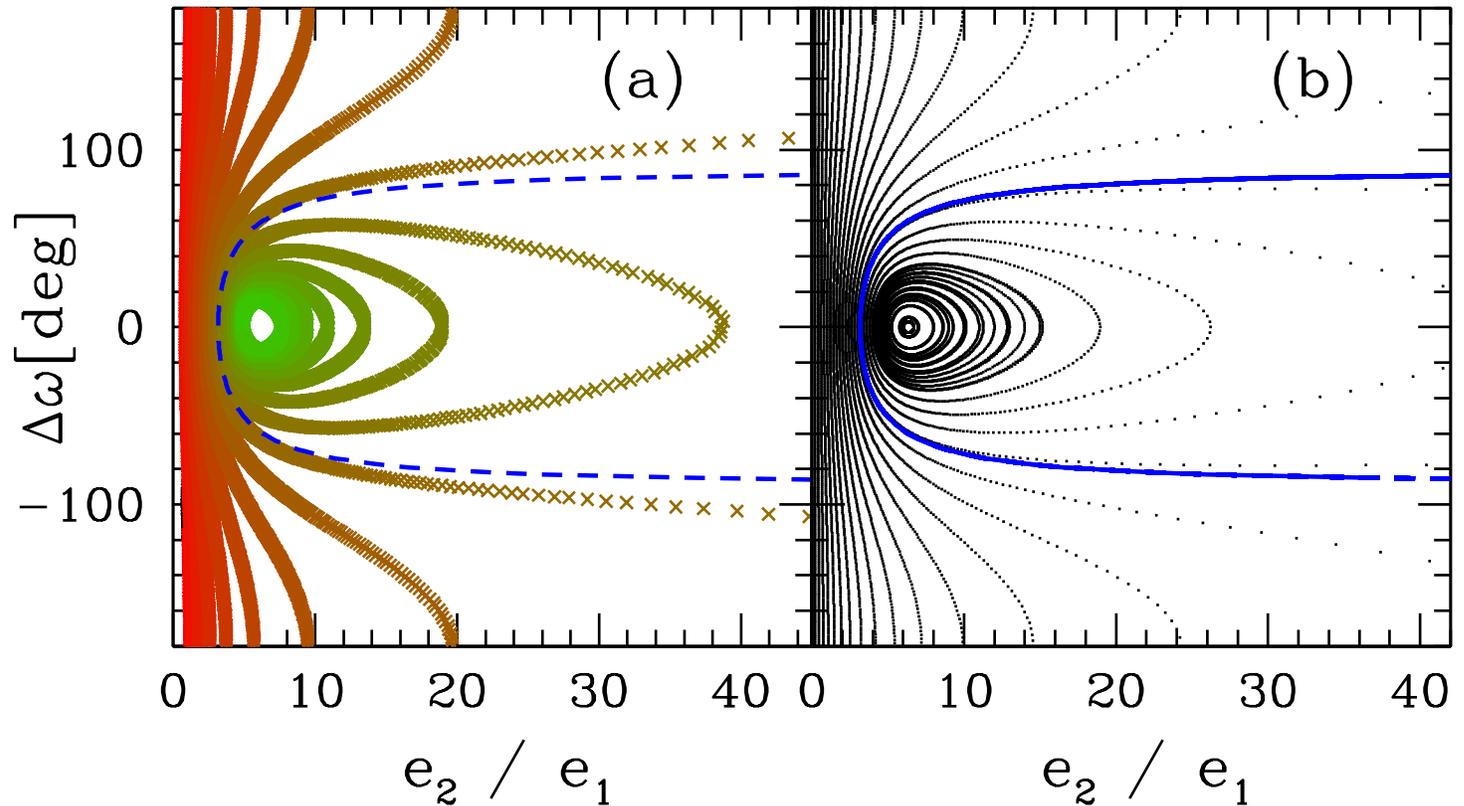}}
\caption{Secular contours drawn in the $\Delta \omega$ -- $e_2/e_1$ space, from a numerical simulation (left) and the classical L-L theory (right).  The numerical simulation started from initially equal eccentricities of 0.01, with $m_1 = 0.05\Mj$ at 1\,AU  and $m_2 = 1\Mj$ at 8\,AU.  Each cross is 1,000 years apart, starting from the circulation regime at $e_2 / e_1 = 1$ and $\Delta \omega_0 = 90$.  In $\sim 0.3\,$Myr, the system evolves past the secular separatrix (dashed curve) and eventually falls in to the small amplitude libration island in $\sim 10\,$Myr.  Similar trajectories derived from the L-L theory are drawn in the right panel for comparison.  \label{contour}} 
\end{figure}

The numerical result of the sample system in Figure~\ref{sample_rigid} is plotted in an $e_2/e_1$ -- $\Delta \omega$ phase diagram in Figure~\ref{contour}~(a).  Figure~\ref{contour}~(b) shows the secular contours computed from the L-L theory for the same system, with varying initial eccentricities and periastron arguments.  The secular separatrix in the Figure~\ref{contour}~(a) and (b) is drawn from the apsidal libration criteria such that 
\begin{equation}
\frac{e_{2,0}}{e_{1,0}} < - \frac{5}{2}\frac{q \alpha^{3/2} \left[ 1 - (1/8) \alpha^2 \right]}{1 - q \alpha^{1/2}} \cos{\Delta \omega_0} \label{down}
\end{equation}
and
\begin{equation}
\frac{e_{2,0}}{e_{1,0}} > \frac{2}{5} \frac{1-q \alpha^{1/2}}{\alpha \left[ 1 - (1/8) \alpha^2 \right]} \frac{1}{\cos{\Delta \omega_0}} \label{up}
\end{equation}
\citep{laughlin02, zhou03}. Here, $e_{1,0}$ and $e_{2,0}$ are the initial eccentricities of the inner and outer planets, respectively, $q\equiv m_1/m_2$, $\alpha \equiv  a_1/a_2$, and $\Delta \omega_0 \equiv \omega_{2,0} - \omega_{1,0}$ is the initial angular offset between the apsidal vectors of the planets.  \citet{zhou03} called  the libration modes defined by Equation~(\ref{down}) and (\ref{up}) ``down-libration'' and ``up-libration'', respectively.  The separatrix in Figure~\ref{contour} corresponds to the up-libration boundary for which the two apsidal vectors are aligned (they are anti-aligned for the down-libration).  

The evolutionary paths of two planets around a single star can be determined solely by their initial conditions, i.e., for each set of initial orbital parameters there is one corresponding secular contour that dictates the evolutionary sequence of the planets.  With the presence of a binary companion, however, the outer planet's eccentricity is secularly excited through the Kozai mechanism, permitting the system to explore different evolutionary modes \footnote{Here we assume that the Kozai oscillation of the inner planet is suppressed by the secular torque applied by the outer planet.  Normally this assumption is valid for a large range of $a_1$ when $m_1 < m_2$. }. 

In Figure~\ref{contour}~(a), the simulation starts at $e_{0,1}/e_{0,2} = 1$ and $\Delta \omega_0 = 60\deg$.  The system is initially placed in the circulation regime so that the two apses precess independently.  However, as $e_2$ starts to evolve significantly through the Kozai mechanism while $e_1$ remaining at a small value, the eccentricity ratio increases, driving the system rightward in the $e_2/e_1$ -- $\Delta\omega$ space.  The system marches on eventually past the secular separatrix, entering the libration regime.  Further increase in $e_2$ brings the system toward progressively smaller libration islands and therefore damps the libration amplitude.  The system maintains a small-amplitude secular libration while $e_2$ is sufficiently large ($\ga 0.1$), until the outer planet returns to a near-circular orbit at the end of its first Kozai cycle.  The system periodically crosses the secular separatrix whenever $e_2$ goes below or above $\sim 0.1$.  Once the system is trapped in the tight apsidal libration regime, the eccentricity ratio is maintained at $\sim 10$, thus $e_1$ also oscillates together with $e_2$ with a smaller amplitude.

\section{Discussion}
The rigid rotation of the planetary orbits and synchronized eccentricity oscillations are frequently observed among our simulations.  The criterion for the up-libration in Equation~\ref{up} can be easily satisfied if the Kozai oscillation amplitude of the outer planet is sufficiently large.  However, the system also needs to maintain its stability for sustainable Kozai oscillations.  One necessary stability requirement is the Hill stability \citep{gladman93}.  The Hill stability needs to be maintained at all times including when $e_2$ is near its maximum.   Another case of instability arises when the initial conditions are chosen such that the system undergoes down-libration (Eq.~\ref{down}) is satisfied.  During down-librations the apsidal vectors are anti-aligned ($\Delta \omega \approx 180\deg$), and $e_2/e_1 \sim 1$ or smaller.  If the down-libration is maintained until the outer planetary orbit embarks on the first Kozai cycle, then $e_1$ is excited to a value larger than $e_2$.  Such an over-excitation leads to a fairly chaotic behavior of the inner orbital eccentricity and will most likely result in  collision of the inner planet with the central star or ejection from the system.

The synchronized eccentricity oscillation mechanism offers a few theoretical implications.  First, in this mechanism the outer planet helps the angular momentum transfer from the companion star to the inner planet.  Propagation of eccentricity disturbance by a stellar passerby through a system of planets has been previously studied by \citet{zakamska04}.  In the case of two planets in a binary, the cyclic propagation of the stellar perturbation through the outer planet can effectively excite the inner planet's eccentricity to oscillate.  This eccentricity propagation mechanism may imply the presence of an undetected secondary planet around known single-planet systems in binaries.  Especially for single planets in wide binaries with the Kozai period too large compared to the estimated age of the system, yet with non-negligible observed eccentricities that are difficult to be explained by other dynamical processes, there is a possibility that there is an undetected secondary planet propagating the eccentricity excitation from the distant companion.  

Another dynamically relevant outcome is the subsequent evolution of the planets after the Kozai mechanism induces instability to the system .  Many systems in our simulations have suffered close encounters between planets that resulted in losing a planet from the system by ejection or collision.  The orbital radius and the  eccentricity of the survived planet can take wide ranges of values, as investigated by planet scattering experiments \citep{juric07, chatterjee07}.  The survived planet may resume the Kozai mechanism, but now the planet's orbital eccentricity is likely to have been already excited to a significantly larger value after ejecting the other planet.  Since the fraction of the time a planetary orbit remains highly eccentric during the Kozai oscillations is strongly dependent on its initial eccentricity \citep{holman97}, single-planet systems in binaries with an episode of early scattering has a much higher probability to be observed at large eccentricities.  For example, known single planets in binaries with very large orbital eccentricities (e.g., HD~80606~b, $e=0.93$ and HD~20782~b, $e=0.92$) strongly suggest the past occurrence of planet ejections in these systems, enabling the eccentricities of the survived planets to oscillate at systematically larger values.


\acknowledgements We would like to thank Ji-Lin Zhou and Fred Adams for beneficial discussions.



\begin{thebibliography}{}
\bibitem[Chatterjee, Ford \& Rasio(2007)]{chatterjee07}
Chatterjee, S., Ford, E. A., \& Rasio, F. A. 2007, arXiv:astro-ph/0703166
\bibitem[Chiang \& Murray(2002)]{chiang02}
Chiang, E. I., \& Murray, N. 2002, \apj, 576, 473
\bibitem[Desidera \& Barbieri(2007)]{desidera07}
Desidera, S., \& Barbieri, M. 2007, \aap, 462, 345
\bibitem[Fabrycky \& Tremaine(2007)]{fabrycky07}
Fabrycky, D., \& Tremaine, S. 2007, arXiv:0705.4285
\bibitem[Ford, Kozinsky \& Rasio(2000)]{ford00}
Ford, E. B., Kozinsky, B., \& Rasio, F. A. 2000, \apj, 535, 585
\bibitem[Gladman(1993)]{gladman93}
Gladman, B. 1993, Icarus, 106, 247
\bibitem[Holman, Touma \& Tremaine(1997)]{holman97}
Holman, M., Touma, J., \& Tremaine, S. 1997, \nat, 386, 254
\bibitem[Innanen et al.(1997)]{innanen97}
Innanen, K. A., Zheng, J. Q., Mikkola, S., \& Valtonen, M. J. 1997, \aj, 113, 1915
\bibitem[Juric \& Tremaine(2007)]{juric07}
Juric, M., \& Tremaine, S. 2007, arXiv:astro-ph/0703160
\bibitem[Kozai(1962)]{kozai62}
Kozai, Y. 1962, \aj, 67, 591
\bibitem[Laughlin, Chambers \& Fischer(2002)]{laughlin02}
Laughlin, G., Chambers, J., \& Fischer, D. 2002, \apj, 579, 455
\bibitem[Murray \& Dermott(1999)]{murray99}
Murray, C. D., \& Dermott, S. F. 1999, Solar System Dynamics (Cambridge University Press)
\bibitem[Raghavan et al.(2006)]{raghavan06}
Raghavan, D., Henry, T., Mason, B. D., Subasavage, J. P., Jao, W., Beaulieu, T. D., \& Hambly, N. C. 2006, \apj, 646, 523
\bibitem[Takeda \& Rasio(2005)]{takeda05}
Takeda, G, \& Rasio, F. A. 2005, \apj, 627, 1001
\bibitem[Wu \& Murray(2003)]{wu03}
Wu, Y, \& Murray, N. 2003, \apj, 589, 605
\bibitem[Veras \& Armitage(2007)]{veras07}
Veras, D., \& Armitage, J. P. 2007, \apj, 661, 1311
\bibitem[Zakamska \& Tremaine(2004)]{zakamska04}
Zakamska, N. L., \& Tremaine, S. 2004, \aj, 128, 869
\bibitem[Zhou \& Sun(2003)]{zhou03}
Zhou, J, \& Sun, Y 2003, \apj, 598, 1290
\end{thebibliography}
\end{document}